\documentclass[aps,prb,amsmath,amssymb,superscriptaddress,twocolumn,color,epsfig,graphicx,bm,floatfix]{revtex4-1}

\usepackage{graphicx}
\usepackage{epstopdf}
\usepackage{tabularx}
\usepackage[super]{nth}
\usepackage[caption=false]{subfig}
\usepackage{amsmath}    
\usepackage{mathtools}
\usepackage{siunitx}
\usepackage{multirow}
\usepackage{array}
\usepackage[version=4]{mhchem}
\newcolumntype{K}[1]{>{\centering\arraybackslash}p{#1}}

\usepackage{verbatim}   
\usepackage{color}      
\usepackage{hyperref}   
\begin{document}

\title{Cubine, a superconducting 2-dimensional copper-bismuth nano sheet} 

\author{Maximilian Amsler}
\affiliation{Department of Materials Science and Engineering, Northwestern University, Evanston, Illinois 60208, USA}
\author{Zhenpeng Yao}
\affiliation{Department of Materials Science and Engineering, Northwestern University, Evanston, Illinois 60208, USA}
\author{Chris Wolverton}
\affiliation{Department of Materials Science and Engineering, Northwestern University, Evanston, Illinois 60208, USA}

\date{\today}

\begin{abstract}
We report on the discovery of a 2-dimensional copper-bismuth nano sheet from \textit{ab initio} calculations, which we call cubine. According to our predictions, single layers of cubine can be isolated from the recently reported high-pressure CuBi bulk material (metastable at ambient conditions) at an energetic cost of merely $\approx 20$~meV/\AA$^2$, comparable to separating single layers of graphene from graphite. Our calculations suggest that cubine has remarkable electronic and electrochemical properties: It is a superconductor with a moderate electron-phonon coupling $\lambda=0.5$, leading to a $T_c$ of $\approx1$~K, and can be readily intercalated with lithium with a high diffusibility, rendering it a promising candidate material as an anode in lithium-ion batteries.  
\end{abstract}

\maketitle

\section{Introduction\label{sec:introduction}}


Since the first successful exfoliating of graphene in 2004,~\cite{Novoselov2004,Yoo2008,Wu2010,Wang2010} intense research has been devoted to the synthesis and characterization of 2D materials. Meanwhile, various 2D materials have been discovered, ranging from graphene analogs like silicene~\cite{takeda_theoretical_1994,guzman-verri_electronic_2007,Cahangirov2009,vogt_silicene:_2012} to phosphorene~\cite{Li2015} and borophene~\cite{amsler_conducting_2013,tang_novel_2007,tang_first-principles_2010,lu_binary_2013},  transition metal oxides,~\cite{Ni2015} transition metal dichalcogenides,~\cite{Xiao2010,Bhandavat2012} and transition metal carbinides/nitrides~\cite{Naguib2012,Er2014}. Due to their diverse properties, 2D materials have been considered for many industrially relevant applications, including  nano electronics, optoelectronics and photonics~\cite{ugeda_giant_2014,qian_quantum_2014,wang_electronics_2012,radisavljevic_mobility_2013,mak_tightly_2013}, but also in the field of energy storage. For metal-ion batteries, materials that can host Li$^+$ or Na$^+$ at high densities are of particular interest, a market currently dominated by carbonaceous (e.g. graphite and hard carbon) and layered materials (e.g. \ce{LiCoO2})~\cite{Peng2016}. 2D materials are promising alternatives since they exhibit compelling ion transport and storage characteristics due to their open 2D channels and high specific surface areas with a large number of active sites.~\cite{Peng2016} Furthermore, since 2D materials only exhibit weak interlayer bonding via van-der-Waals (vdW) forces, unfavorable, large changes in their volumes during the intercalation and de-intercalation of lithium ions are alleviated. 2D materials can also serve as functional substrates for incorporating active materials to improve the electrical and ionic conductivity of electrodes~\cite{Wu2010,Yang2010}, and are promising candidates for applications in nanostructured supercapacitors.~\cite{YuZ2015}


Theoretical efforts to identify potential new 2D materials have been aimed at screening large databases of crystalline materials for low-dimensional structural patterns~\cite{lebegue_two-dimensional_2013,ashton_topology-scaling_2017}. Commonly, simple geometrical criteria such as large interlayer spacings are used to identify 2D units that could be separated through mechanical exfoliation. The recently discovered Cu--Bi intermetallics have attracted considerable attention due to such low-dimensional structural features, including 1D channels and 2D voids~\cite{Clarke2016,guo_weak_2017}. Despite the strong immiscibility at ambient pressure~\cite{Chang1997}, Cu--Bi binary compounds have been known to form at high pressures since the early 1960s, exhibiting superconducting properties with a $T_\mathrm{c}$ between 1.33 and $\SI{1.40}{\kelvin}$~\cite{Matthias1961}. In 2016, a \ce{Cu11Bi7} compound was reported to form at moderate pressures, which crystallizes in a hexagonal structure with a $T_\mathrm{c}$ of $\SI{1.36}{\kelvin}$~\cite{Clarke2016}. Recently, another high-pressure phase in the Cu--Bi system at the equiatomic composition, CuBi, was discovered with an orthorhombic unit cell~\cite{guo_weak_2017}. Despite the dissimilarities in their crystal structures, both \ce{Cu11Bi7} and  CuBi share a common structural motif, namely the formation of voids to host the electron lone pairs (ELP) that form in the vicinity of the bismuth atoms. In contrast to \ce{Cu11Bi7}, which exhibits channels laced by ELP, CuBi forms a layered structure of alternating Cu, Bi, and ELP sheets. The formation of this low-density phase is the more surprising since such materials strongly violate the common perception that high-pressure phases have high densities with short, strong interatomic bonds.


Here we investigate the \ce{CuBi} compound using \textit{ab initio} methods to unravel the unique structural features of CuBi and reveal its 2-dimensional character. Based on our calculations, layers of CuBi are held together by weak vdW forces, and we demonstrate that single nano-sheets of CuBi can be separated from the bulk material at an extremely low cost in energy, one that is comparable to values required to exfoliate graphene from graphite. The dynamical stability of CuBi nano sheets show that they are viable as metastable phases, and could be potentially used as a new class of building blocks in 2D heterostructures with compelling physical properties: they are  superconducting with a $T_\mathrm{c}\approx 1$~K, and could be used to intercalate Li atoms at low voltages, and hence make an interesting new anode for Li-ion batteries.


\section{Results and discussion\label{sec:results}}
\begin{figure}
\subfloat[Bulk CuBi\label{sfig:structa}]{%
\includegraphics[width=0.4\columnwidth]{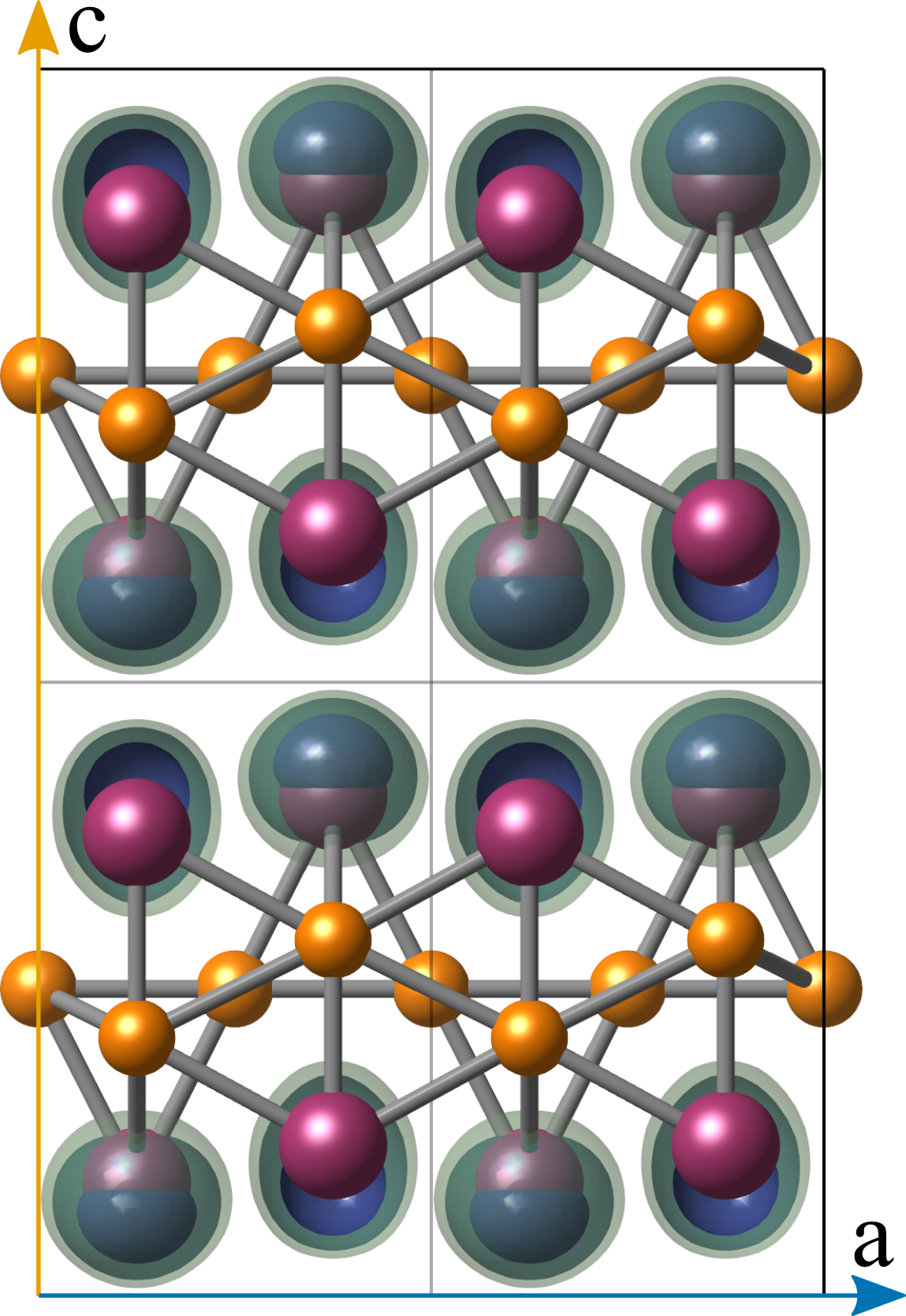}
}\hfill
\subfloat[Single sheet of cubine\label{sfig:structb}]{%
\includegraphics[width=0.56\columnwidth]{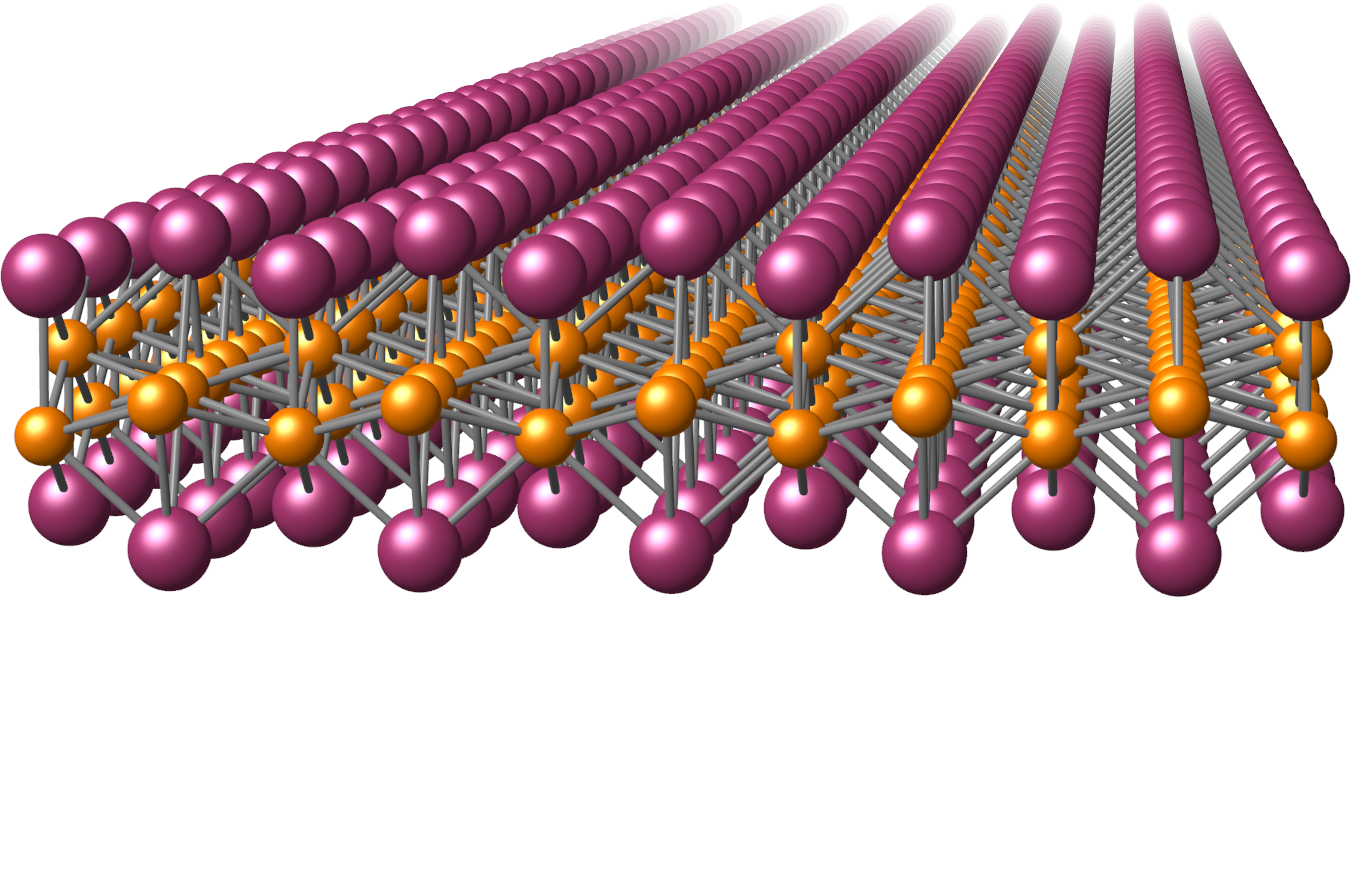}
\vspace{1cm}
}
\caption{ 
Panel (a) shows the bulk CuBi structure along the b-axis together with the isosurfaces of the ELF drawn at values of 0.9, 0.8 and 0.7. A single sheet of cubine, exfoliated from CuBi, is shown in panel (b) along the a-axis. Cu and Bi atoms are denoted by  orange (small) and purple (large) spheres.\label{fig:struct}}

\end{figure}

The thermodynamic stability of a solid state compound is governed by the Gibbs free energy, $G=E+pV-TS$, where the phase (or combination of phases) with the lowest value is the stable state at a given pressure $p$ and temperature $T$. At high pressure, the $pV$ term becomes increasingly dominant, and therefore high density polymorphs will be favored and a strong penalty is expected for layered, vdW bonded materials. The most prominent example for this behavior is carbon, where graphite, essentially consisting of layered graphene sheets and stable at ambient conditions, transforms into the dense cubic diamond structure at high pressures~\cite{naka_direct_1976}. Similarly, \ce{N_2}, a quasi-0D material at ambient pressure, turns into a polymeric phase at above 120~GPa~\cite{eremets_single-bonded_2004}, following the trend that the dimensionality and density of accessible polymorphs increases with pressure.

The recently discovered high-pressure CuBi material however poses a puzzling exception to this general rule, since pieces of evidence indicate that CuBi is in fact a layered material with a strong 2D character. CuBi crystallizes in an orthorhombic cell with \textit{Pmma} symmetry at above 3.2~GPa., and its structure is shown in panel (a) of Figure~\ref{fig:struct}  along the (010) direction. Two Cu atoms occupy the 2d and 2e sites at the Wyckoff positions (0,$\frac{1}{2}$,$\frac{1}{2}$) and ($\frac{1}{4}$,0,0.579), respectively, and two Bi atoms are on the 2e and 2f sites at  ($\frac{1}{4}$,0,0.245) and  ($\frac{1}{4}$,$\frac{1}{2}$,0.815), respectively. According to our calculations with the  PBE functional, the lattice constants at 0~GPa are $a=5.204$~\AA,  $a=4.263$~\AA, and $a=8.131$~\AA\, (see Table~\ref{tab:exfoliation}). While the values for $a$ and $b$ are in excellent agreement with experiments with errors of less than 0.5~\%, the $c$ vector is overestimated by roughly 3.3~\%. A very similar behavior is observed in graphite~\cite{rego_comparative_2015}, where the interlayer distance is also severely overestimated, a first indication of the 2D character of CuBi. Furthermore, in both systems, this overestimation can be corrected by employing the LDA functional, or by using dispersion corrected PBE functionals as shown in Table~\ref{tab:exfoliation}. These findings clearly indicate that the interlayer interaction in CuBi are vdW mediated, a characteristic feature of layered 2D materials.

We can also see that CuBi is a layered material from a simple geometric argument. The main structural motif consists of layers of buckled triangular sheet composed purely of Cu atoms in the \textit{a--b} plane, with two unique, metallic Cu--Cu bonds. Along the \textit{b}-direction the bond length is 2.601~\AA, whereas along the \textit{a}-direction it is 2.580~\AA. Bi atoms are attached to both sides of these Cu sheets along the \textit{c}-direction, with a shortest Cu--Bi distance of 2.720~\AA. Two further symmetrically inequivalent bonds are formed between Cu and Bi at distances of 2.966~\AA\, and 3.245~\AA, respectively. No bonds are formed between the Bi atoms, and all Bi--Bi distances are significantly larger than the sum of the covalent radii. In particular, the distance between the layers of Bi is 3.98~\AA, clearly indicating that the bonding between the layers is neither covalent nor metallic. Furthermore, the electron localization function in Figure~\ref{fig:struct} shows no Bi--Bi bonds, but instead reveals that stereochemically active ELP are interlaced between the Bi layers.

\begin{figure*}[!htbp]
\centering
\subfloat[\label{fig:bulkph}]{\includegraphics[width=1.2\columnwidth]{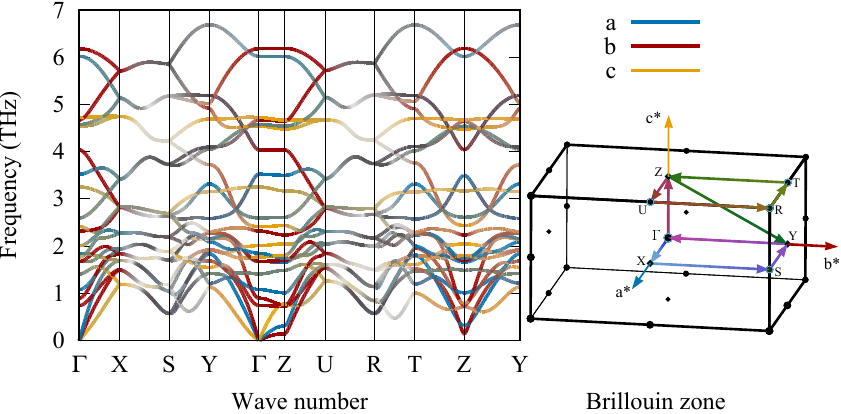}}\hfill
\subfloat[\label{fig:vols}]{\includegraphics[width=0.85\columnwidth,angle=0]{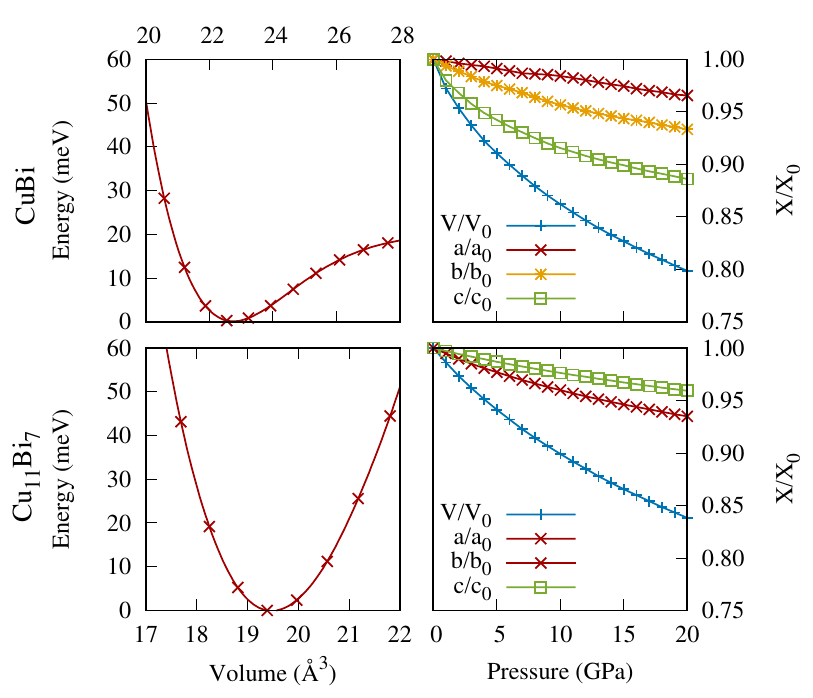}}
\caption{(a) The phonon dispersion of the bulk material is shown along a path in the first Brillouin zone. The colors indicate the contribution of the vibrational eigenmodes along the three crystal lattices a, b and c. (b) The left panels show the energy per atom as a function of volume, top for \ce{CuBi} and bottom for \ce{Cu11Bi7}. The right panels show the evolution of the lattice parameters and the unit cell volume normalized to the ambient pressure values, for \ce{CuBi} and \ce{Cu11Bi7} at the top and bottom, respectively.}
\end{figure*}

A detailed analysis of the lattice vibrations provides further evidence for the 2D character of CuBi. Figure~\ref{fig:bulkph} shows the phonon dispersion together with the  first Brillouin zone. We identify three specific features in the phonon band structure that are characteristic for 2D materials. First, the lowest energy acoustic branch (yellow) shows a quadratic behavior as $q\rightarrow \Gamma$ along the in-plane directions $X-\Gamma$ and $Y-\Gamma$, typical for 2D materials that exhibit a single flexural phonon branch with quadratic dispersion~\cite{carrete_physically_2016_m}. Second, the very flat bands along the $\Gamma-Z$ with essentially zero dispersion can be directly attributed to extremely weak interactions between the layers. Third, the two lowest energy branches along $\Gamma-Z$ are dominated by collective in-plane vibrations of the atoms  in the $a$ and $b$ directions. This behavior is shown by the color coding based on the magnitude of the projected vibrational eigendisplacements along the three lattice vector components.

The anisotropic compressibility and strong anharmonicity of CuBi is the fourth  evidence for its 2D character. Figure~\ref{fig:vols} compares the structural parameters upon compression and expansion of CuBi with the recently discovered \ce{Cu11Bi7} phase~\cite{Clarke2016}. While the latter exhibits a roughly  quadratic dependence of the energy on the volume, CuBi behaves strongly anharmonically, similar to graphite~\cite{bahmann_crossed_2013}. The right panels in Figure~\ref{fig:vols} show the pressure dependence of the normalized unit cell volumes and the cell parameters upon compression of CuBi and \ce{Cu11Bi7}. In contrast to \ce{Cu11Bi7}, the change in the lattice vectors of CuBi is highly anisotropic, with a higher compressibility along the $c$-direction compared to $a$ and $b$.

\begin{figure*}[!htbp]
\centering
\subfloat[\label{fig:2Dph}]{\includegraphics[width=1\columnwidth]{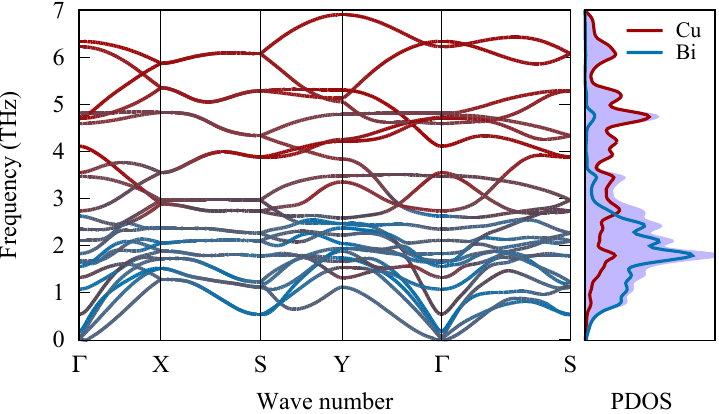}}\hfill
\subfloat[\label{fig:2Del}]{\includegraphics[width=1\columnwidth]{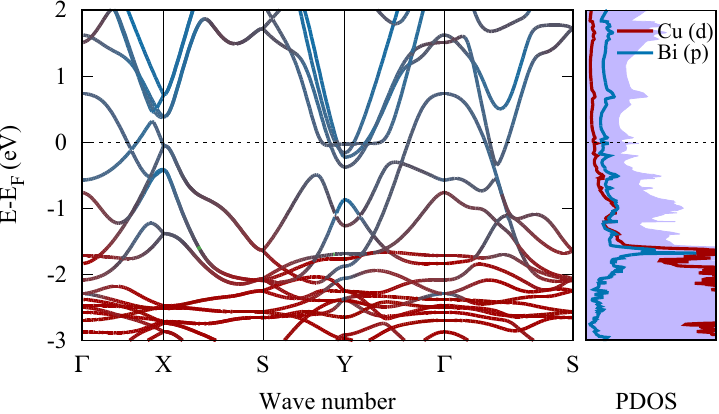}}
\caption{(a) The phonon band structure of cubine, the 2D isolated sheet of CuBi, together with the partial density of states (PDOS), indicating that no imaginary modes arise within the complete 2D Brillouin zone. The red and blue color coding denotes the modes dominated by Cu and Bi vibrations, respectively. The shaded area in the PDOS represents the total density of states. (b) The electronic band structure  of cubine, where the colors indicate the band character projected onto the Cu and Bi atoms in red and blue, respectively. The shaded area in the right panel represents the total density of states, whereas the red and blue lines denote the Cu $3d$ and Bi $6p$ contributions, respectively. 
}

\end{figure*}

Based on all these properties supporting the 2D character of CuBi, we investigated if a single layer of CuBi, so-called cubine, is in fact viable. The two necessary criteria that need to be satisfied are that cubine is dynamically stable, and that there is only a moderate-to-weak energy penalty for isolating single sheets. Figure~\ref{fig:2Dph} shows the phonon dispersion in the 2D Brillouin zone. No imaginary phonons are present, indicating that single sheets of cubine are indeed metastable. The color coding of the bands and the partial density of states show that two frequency regions can be clearly distinguished. The low frequency regime below $\approx3$~THz is strongly dominated by vibrations of the heavy Bi atoms, whereas the lighter Cu atoms contribute to the high energy phonons.

To assess if cubine can be isolated, we computed the interlayer binding energy $E_i$ between the individual sheets according to $E_i=\frac{E_\text{sheet}-E_\text{bulk}}{A}$, where $E_\text{bulk}$ is the total bulk energy, and $A$ is the area spanned by the in-plane lattice vectors. $E_\text{slab}$ is the energy of a single layer of cubine, evaluated by introducing a vacuum layer of 12~\AA \, between the individual sheets. The PBE functional results in an extremely low value in $E_i$ of merely 4~meV/\AA$^2$. However, since semilocal DFT does not account for vdW interactions, the real interlayer energy might be significantly higher. Indeed, the values of $E_i$ by using functionals that (empirically) take into account dispersion energies, are overall higher than PBE and lie consistently in the range of $\approx 15-30$~meV/\AA$^2$ (see Table~\ref{tab:exfoliation}). These values are close to the interlayer energies of graphite and other 2D materials~\cite{bjorkman_van_2012,lebegue_two-dimensional_2013,ashton_topology-scaling_2017}, and therefore strongly support the potential of exfoliating cubine from bulk in analogy to graphene~\cite{Novoselov2004}.

In order to compare with available literature data, we screened the large 2D materials database at \href{http://www.materialsweb.org}{www.materialsweb.org} for structures similar to cubine, but none of the 15 binary $AB$ prototype structures~\cite{ashton_topology-scaling_2017} matches its space group symmetry. However, two compounds have similar structural motifs of a core Cu sheet between layers of a second atomic species, namely \ce{CuTe} and \ce{CuBr}. In contrast to cubine,  \ce{CuBr} has a completely planar Cu layer (space group $P4/nmm$), while \ce{CuTe} exhibits a slightly different stacking geometry  (space group $Pmmn$). Hence, the structure of cubine (space group $Pmma$) is presumably novel and we propose it to be added to the database as the \nth{16} member of $AB$ prototypes.

\begin{table}[!htbp]
\caption{\label{tab:exfoliation} The lattice parameters (in \AA) are give for CuBi and graphite in their crystalline phase (columns 3-5) and as isolated layers of cubine and graphene (columns 6 and 7) using various exchange-correlation functionals. The last columns contains the interlayer energies $E_i$ in meV/\AA$^2$. The following abbreviations were used for the various dispersion corrected PBE functionals: Grimme D2 (D2)~\cite{grimme_semiempirical_2006}, Grimme D3 without (D3)~\cite{grimme_consistent_2010} and with (D3-BJ)~\cite{grimme_effect_2011}  Becke-Johnson dampening, Tkatchenko-Scheffler method with iterative Hirshfeld partitioning (TS-HP)~\cite{bucko_extending_2014}, and Steinmann's density-dependent dispersion energy correction dDsC~\cite{steinmann_comprehensive_2011}.}
\begin{ruledtabular}
\begin{tabular}{l c|c  c  c | c c | c}
\multicolumn{2}{c}{}     & \multicolumn{3}{c}{Bulk} &  \multicolumn{2}{c}{Sheet}\\
& Method &  $a$  & $b$ & $c$  & $a$  &  $b$ &   $E_i$   \\
 \hline
Cubine & PBE      & 5.204 & 4.263 & 8.131  &  5.197 &   4.272  & 4  \\
       & LDA      & 5.085 & 4.125 & 7.817  &  5.055 &   4.145  & 21 \\
       & D2       & 5.161 & 4.235 & 7.916  &  5.164 &   4.221  & 31 \\
       & D3       & 5.066 & 4.250 & 7.901  &  5.054 &   4.255  & 20 \\
       & D3-BJ    & 5.151 & 4.181 & 7.743  &  5.095 &   4.209  & 24 \\
       & TS-HP    & 5.061 & 4.243 & 7.907  &  5.056 &   4.245  & 20 \\ 
       & dDsC     & 5.164 & 4.210 & 7.895  &  5.130 &   4.227  & 14 \\
       & Exp.~\cite{guo_weak_2017}     & 5.207 & 4.242 & 7.876  &  --    &   --     & -- \\
\hline
Graphene& PBE~\cite{rego_comparative_2015}  & 2.466 & &8.755&  -- &    &   1       \\
& PBE~\cite{jiang_first_2009}  & -- &  & -- & 2.465  &     &          -- \\
& LDA~\cite{lebegue_cohesive_2010}          & 2.460 & &6.660& --  &      &   9       \\ 
& LDA~\cite{sahin_monolayer_2009}          & -- & & -- &  2.460 &      &    --    \\ 
& D2~\cite{rego_comparative_2015}           & 2.461 & &6.444&   --&      &   21      \\
& D3~\cite{rego_comparative_2015}           & 2.464 & &6.965&   --&      &   18      \\
& Exp.~\cite{zacharia_interlayer_2004}      & 2.46  & &  -- &   --&      &   $20\pm2$  \\ 
& Exp.~\cite{benedict_microscopic_1998}     & 2.46  & &6.70 &   --&      &   13      \\
\end{tabular}
\end{ruledtabular}
\end{table}


\begin{figure}
\includegraphics[width=0.9\columnwidth]{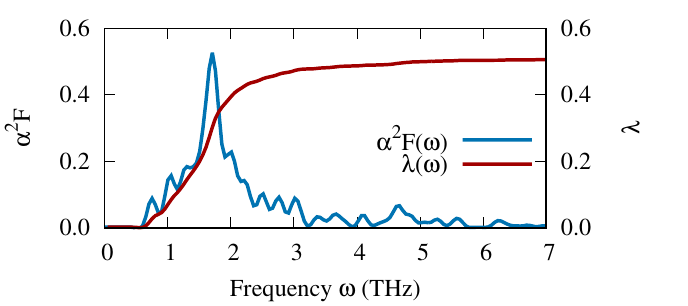}
\caption{The Eliashberg spectral function $\alpha^2F(\omega)$ of cubine together with the integrated electron-phonon coupling parameter $\lambda(\omega)$ as a function of the phonon frequency $\omega$.\label{fig:elphon}}
\label{}
\end{figure}

Similar to its bulk counterpart, cubine is metallic as illustrated by the band structure in Figure~\ref{fig:2Del} with a rather low density of states at the Fermi level. As expected, the strongest contribution at the Fermi level stem from the Cu $3d$ and Bi $6p$ states. The 2 Bi $6s$ electrons of the stereochemically active ELP are buried deep below the valence bands. Since both crystalline bulk (3D) compounds CuBi and \ce{Cu11Bi7} are superconductors, we computed the electron-phonon coupling properties. Figure.~\ref{fig:elphon} shows the Eliashberg spectral function $\alpha^2F(\omega)$ together with the integrated electron-phonon coupling parameter $\lambda(\omega)$. $\alpha^2F(\omega)$ exhibits a strong peak around 2~THz, and  $\lambda(\omega)$ increases sharply in the range of 0-2~THz. A comparison with the PDOS in Figure~\ref{fig:2Dph} reveals that these low energy phonon modes stem primarily from the vibration of the heavy Bi atoms. The resulting coupling strength of $\lambda=0.50$ is rather weak, and considerably lower than for the bulk material, $\lambda=0.67$. Depending on the value of the empirical Coulomb pseudopotential $\mu^*$, the predicted superconducting temperature ranges from 0.6-1.0~K. Slightly lower values were obtained when employing the LDA exchange correlation functional, as indicated by the results summarized in Table~\ref{tab:Tc}.

\begin{table}[]
\caption{Electron-phonon coupling parameters for both the PBE and the LDA exchange correlation functionals, and the resulting superconducting transition temperatures $T_c$ for different values of $\mu^*$\label{tab:Tc}}
\begin{ruledtabular}
\begin{tabular}{l|l c c c c |c c}
 & $\mu^*=$& 0.10 & 0.11  & 0.12 & 0.13 & $\lambda$ & $\omega_\text{log}$ (K)  \\
 \hline
 {PBE}    & $T_c =$ & 1.0~K  & 0.8~K  & 0.7~K  & 0.6~K   &  0.50     &    80  \\
\hline
 {LDA}    & $T_c =$ & 0.7~K  & 0.6~K  & 0.5~K  & 0.4~K   &  0.45     &    86  
\end{tabular}
\end{ruledtabular}
\end{table}


\begin{figure}[!htbp]
\centering
\includegraphics[width=0.8\columnwidth]{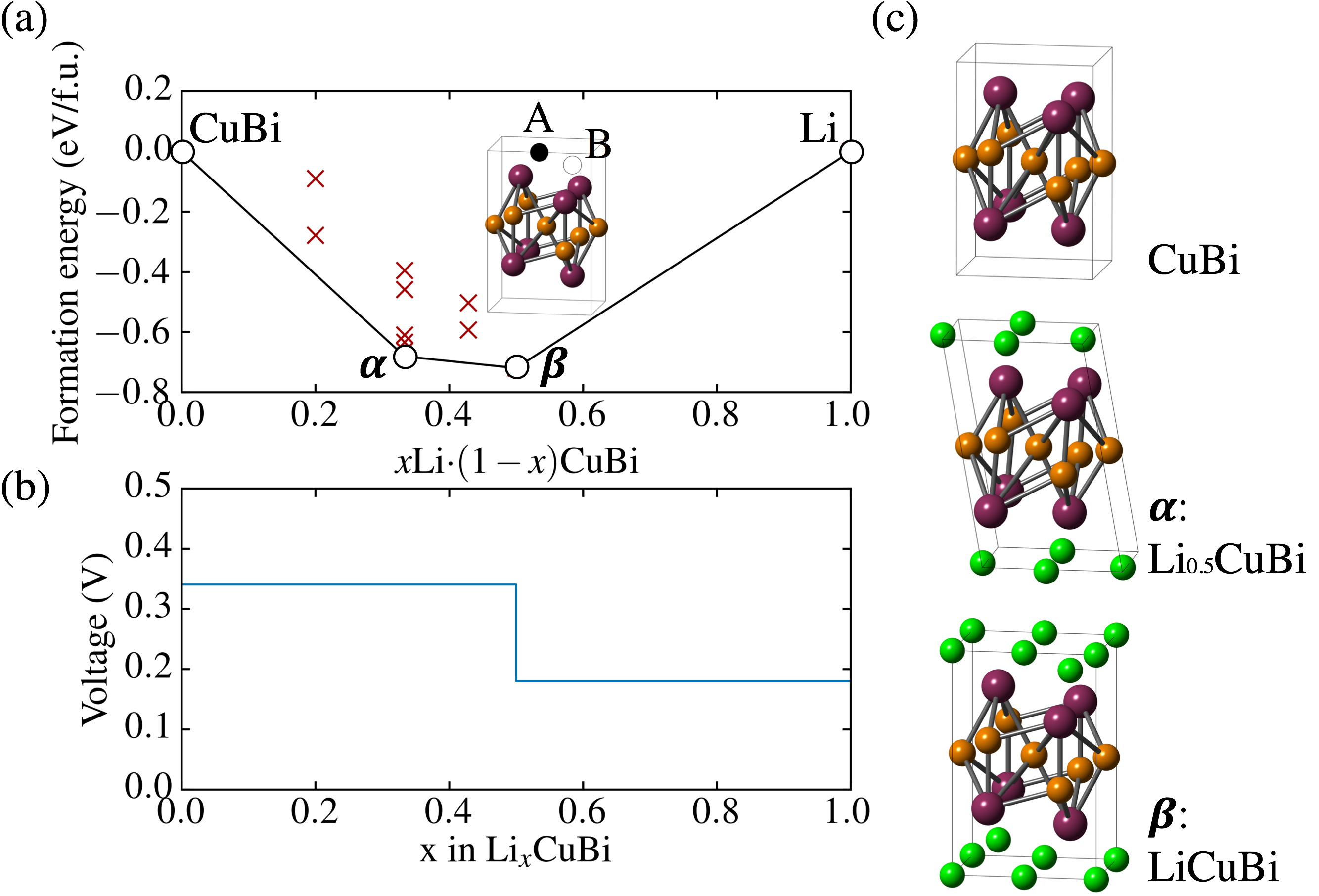}
\caption{(a) CuBi-Li convex hull of stability as a function of lithium concentration, and (b) the corresponding voltage profile. (c) Predicted structures of the intermediate phases.
\label{fig:liocc}}
\end{figure}

The very low interlayer binding energy and the large spacing of 3.98~\AA\, (graphite: 3.35~\AA) between the sheets renders CuBi an attractive candidate as a metal intercalation electrode. There are two symmetrically distinct interstitial sites per cell (inset of panel (a), Figure~\ref{fig:liocc}) with each twofold degeneracy between the cubine layers that can host four guest atoms in total. By fully occupying these sites with Li, Na, Mg and Al, capacities of 96.46 mAh/g, 92.45 mAh/g, 184.27 mAh/g and 274.51 mAh/g could be achieved, respectively. To further study the charge/discharge properties, we will for now only focus on the lithiation electrochemistry of CuBi. There are two thermodynamically stable intermediate phases on the CuBi--Li convex hull, as shown in panel (a) of Figure~\ref{fig:liocc}: Li$_{0.5}$CuBi and LiCuBi. The corresponding voltage profile is shown in panel (b), with values in a fairly low range of 0.18-0.34~V, indicating that this material is more suitable as an anode material. During the whole lithiation process, the CuBi backbone remains completely intact with only slight sliding of the cubine layers around the concentration Li$_{0.5}$CuBi (see panel (c) in Figure~\ref{fig:liocc}).


%

\begin{figure}[!htbp]
\centering
\includegraphics[width=0.8\columnwidth]{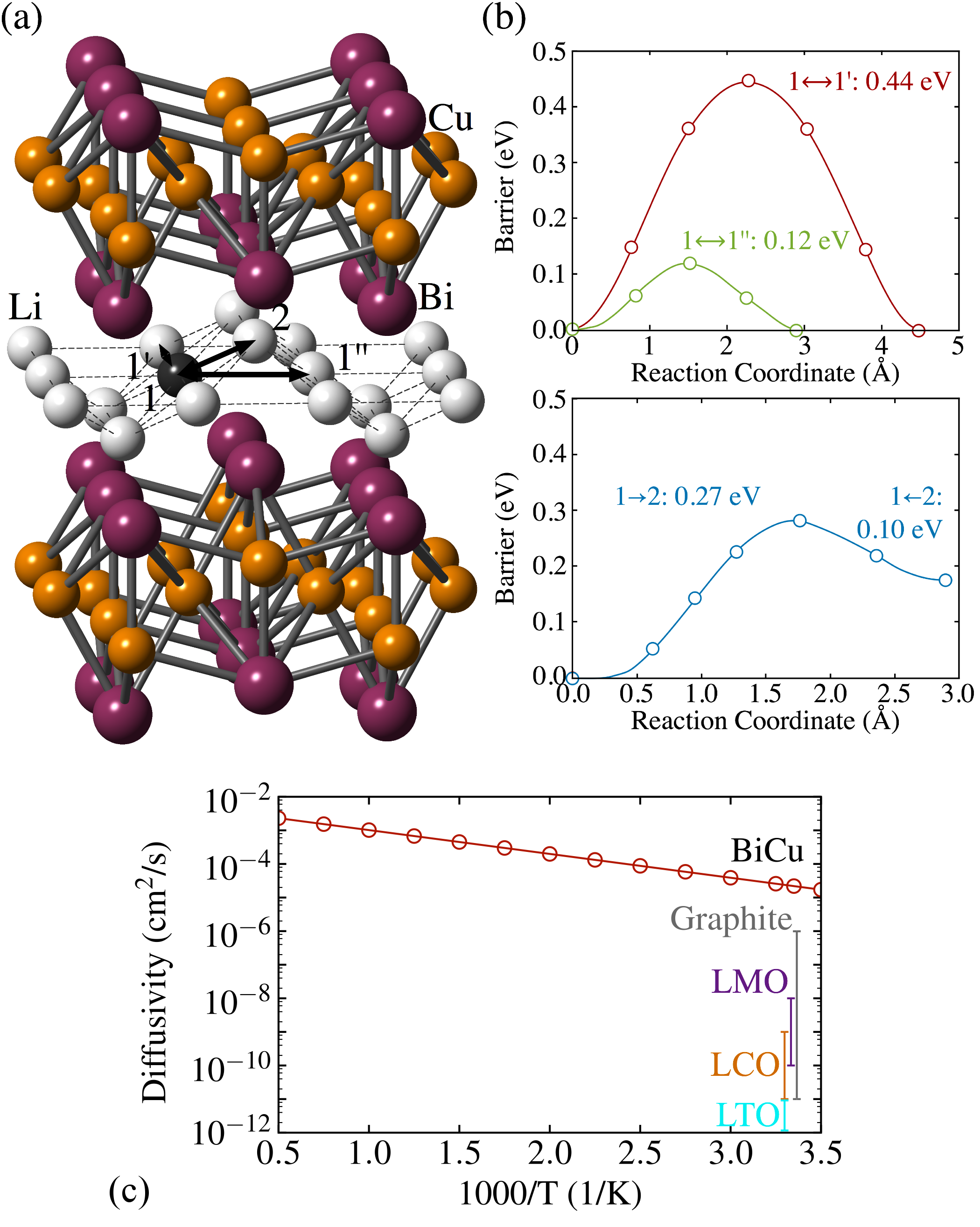}
\caption{(a) Lithium ion diffusion network through the interlayer space of CuBi. (b) Kinetic barriers calculated along geometrically distinct diffusion paths. (c) Calculated lithium ion diffusivity as a function of temperature (attempt frequency: $\nu = 10^{13}$~s$^{-1}$), compared to state-of-the-art electrode materials: Graphite, LCO (LiCoO$_2$), LMO (LiMn$_2$O$_4$), and LTO (Li$_4$Ti$_5$O$_{12}$) with experimental data adopted from Refs.~\onlinecite{Park2010,Nitta2015}.
\label{fig:diffusion}}
\end{figure}

Furthermore, the interlayer diffusion of the Li atom was assessed by computing the barriers along the geometrically distinct paths between the cubine layers (Figure~\ref{fig:diffusion}). The lowest among all barriers was found for the transition between the symmetrically equivalent 1 and 1'' sites with a value of merely 0.12~eV. With all transition barriers at hand, a kinetic Monte Carlo simulation was carried out to estimate the lithium ion diffusivity $D$. The values of $D$ as a function of temperature is shown in Figure~\ref{fig:diffusion}c, calculated using a typical attempt frequency of $\nu=10^{13}$~s$^{-1}$.  We find a superior Li diffusivity of $3.8\times10^{-5}$~cm$^2$/s at room temperature, more than one order of magnitude higher than all state-of-the-art electrodes. For comparison, anodes like graphite or \ce{Li4Ti5O12} (LTO) exhibit values of $D=10^{-6}\sim10^{-11}$~cm$^2$/s and $10^{-11}\sim10^{-12}$~cm$^2$/s, respectively, while  cathodes like \ce{LiCoO2} (LCO) or \ce{LiMn2O4} (LMO) have diffusivities in the range of $D=10^{-8}\sim10^{-10}$~cm$^2$/s and $10^{-9}\sim10^{-11}$~cm$^2$/s), respectively.~\cite{Park2010,Nitta2015} Based on these results, CuBi fulfills all necessary requirements as a promising intercalation-type anode material in rechargeable lithium ion batteries. In fact, with its remarkable Li conductivity at relatively low cost, CuBi is one of the most promising candidates for high-power metal-ion batteries in the electronic vehicle industry, which demand electrodes with ultrafast charging/discharging capabilities.


%

\section{Conclusions}
In summary, we report on the discovery of a novel quasi-2D sheet, cubine. We present many pieces of compelling evidence  to demonstrate that CuBi is indeed a 2D material with weak vdW interactions between single cubine layers. Since the interlayer energy is comparable to graphene, we predict that single sheets of cubine can be exfoliated from CuBi using similar techniques used to isolate graphene from graphite. As a single sheet, cubine can readily serve as a building block for heterostructured 2D materials. Cubine is metallic with a moderate electron-phonon coupling, leading to superconducting transition temperatures in the range of around $0.6-1$~K. Furthermore, we demonstrate that CuBi can be readily intercalated with lithium with a high ion diffusivity, rendering it a promising candidate for energy storage applications.

\section{Acknowledgments}
M.A. and C.W. acknowledge support from the Novartis Universit{\"a}t Basel Excellence Scholarship for Life Sciences, the Swiss National Science Foundation (P300P2-158407), and DOE (DE-FG02-07ER46433). Z.Y. (the Li intercalation and conductivity calculations) was supported as part of the Center for Electrochemical Energy Science (CEES), an Energy Frontier Research Center funded by the U.S. Department of Energy, Office of the Science, Basic Energy Science under award number DE-AC02–06CH11. We gratefully acknowledge the computing resources from: the Swiss National Supercomputing Center in Lugano (project s700), the Extreme Science and Engineering Discovery Environment (XSEDE) (which is supported by National Science Foundation grant number OCI-1053575), the Bridges system at the Pittsburgh Supercomputing Center (PSC) (which is supported by NSF award number ACI-1445606), the Quest high performance computing facility at Northwestern University, the National Energy Research Scientific Computing Center (DOE: DE-AC02-05CH11231), and Blues, a high-performance computing cluster operated by the Laboratory Computing Resource Center at Argonne National Laboratory. 
\section{Method\label{sec:method}}

Density functional theory (DFT) calculations were carried out within the projector augmented wave~(PAW) formalism~\cite{PAW-Blochl-1994} as implemented in the \texttt{VASP}~\cite{VASP-Kresse-1995,VASP-Kresse-1996,VASP-Kresse-1999} package together with the Perdew-Burke-Ernzerhof (PBE) approximation~\cite{Perdew-PBE-1996} to the exchange correlation potential. A plane-wave cutoff energy of 400~eV was used with a sufficiently dense k-point mesh to ensure a convergence of the total energy to within 1~meV/atom. Both the atomic and cell parameters were simultaneously relaxed until the maximal force components were less than 4~meV/\AA \, and stresses less than 0.01~GPa.

Phonon calculations were carried out with the frozen phonon approach as implemented in the \texttt{PHONOPY} package~\cite{phonopy}, using sufficiently large supercells for converged thermal properties. To compute the vibrational contribution to the density of states a dense mesh of $30\times30\times30$ was used to sample the irreducible Brillouin zone.

The \texttt{Quantum Espresso} package~\cite{espresso} was used to compute the phonon-mediated superconducting properties. We used norm conserving FHI pseudopotentials and a plane-wave cutoff energy of 150~Ry. The Allan-Dynes modified McMillan's approximation of the Eliashberg equation~\cite{Allen_1975} was used to estimate the superconducting temperature: 
\begin{equation}\label{eq:mcmillan}
  T_\text{c}=\frac{\omega_\text{log}}{1.2}\exp\left[-\frac{1.04(1+\lambda)}{\lambda-\mu^{*}(1+0.62\lambda)}\right]
\end{equation}
where $\mu^{*}$ is the Coulomb pseudopotential, $\lambda$ is the overall electron-phonon coupling strength computed from the frequency dependent Eliashberg spectral function $\alpha^2F(\omega)$, and $\omega_\text{log}$ is the logarithmic average phonon frequency. A $4\times 4\times 1$ $q$-mesh was used together with a denser $32\times 32\times 1$ $k$-mesh for the 2D \textit{Pmma} structure, resulting in  a well converged superconducting transition temperature $T_{\rm c}$. A value of $\mu^*=0.10-0.13$ was employed, which was shown to give $T_{\rm c}$'s in excellent agreement with experimental results for the \ce{Cu11Bi7} superconductor~\cite{Clarke2016}.


To simulate the lithiation process of CuBi, the conventional cell was used to explore geometrically distinct potential Li sites between the cubine layers. All symmetrically inequivalent occupation  of these sites were used as input structures for local relaxations, resulting in energies as a function of Li/vacancy ratios, i.e. the Li concentrations in Li$_x$CuBi. The formation energies were evaluated according to the following reaction: CuBi+$x$Li$\rightarrow$Li$_x$CuBi. The lithiation convex hull was constructed from the lowest formation energies at every composition, and only those intermediate phases were considered which lie on the hull. The average lithiation/delithiation reaction voltage relative to Li/Li$^+$ is given by the negative of the reaction free energy per Li:~\cite{Aydinol1997}
\begin{equation}\label{}
  \bar{V}=\frac{\Delta G_\text{f}}{F\Delta N_{\text{Li}}}
\end{equation}
where $F$ is the Faraday constant, $\Delta N_{\text{Li}}$ is the amount of Li added/removed and $\Delta G_\text{f}$ is the (molar) change in free energy of the reaction. The enthalpic ($p$V$_m$) contribution to $G$ is of the order of 10~$\mu$eV per Li at atmospheric pressure and can be safely ignored, while the entropic contribution to the voltage was assumed to cancel each other out and was therefore neglected.~\cite{Chan2012} Hence, the total DFT internal energies ($E$) were used to approximate $G$:
\begin{equation}\label{eq:delta}
  \Delta G\approx \Delta E=E(\text{Li}_x\text{CuBi})-E(\text{CuBi})-xE(\text{Li}_\text{metal})
\end{equation}
where $E(\text{Li}_x\text{CuBi})$,  $E(\text{CuBi})$  and $E(\text{Li}_\text{metal})$ are the total energies of Li$_x$BiCu, BiCu, and elemental Li, respectively. The lithium ion diffusivity was studied using Kinetic Monte Carlo (KMC) simulations using the Materials Interface package \texttt{MINT}~\cite{MINT} based on barriers computed through the climbing image Nudged Elastic Band ($c$NEB) approach as implemented in the \texttt{VASP TST} package~\cite{Henkelman2000,Michel2014,Li2016}. 



\providecommand{\noopsort}[1]{}\providecommand{\singleletter}[1]{#1}%

\end{document}